\documentclass[twocolumn,showpacs,preprintnumbers,amsmath,amssymb]{revtex4}
\usepackage{graphicx}% Include figure files
\usepackage{dcolumn}% Align table columns on decimal point
\usepackage{bm}% bold math

\newcommand{\half}{{\scriptstyle{\frac{1}{2}}}}
\newcommand{\const}{\mathop{\rm const}\nolimits}
\def\bA{{\bf A}}
\def\su{{\rm su}}
\def\p{{\partial}}
\def\br{{\bf r}}
\def\bk{{\bf k}}
\def\bJ{{\bf J}}

\def\bp{{\bf p}}

\def\bu{{\bf u}}

\def\bv{{\bf v}}
\def\bw{{\bf w}}
\def\bq{{\bf q}}
\def\bN{{\bf N}}
\def\grad{{\bm{\nabla}}}
\newcommand{\rot}{\mathbf{rot}}
\def\bE{{\bf E}}
\def\bB{{\bf B}}
\def\bH{{\bf H}}
\def\bD{{\bf D}}

\begin{document}

\preprint{cond-mat/0509636}

\title{Fermat Principle for spinning light}

\author{C. Duval}
\affiliation{
Centre de Physique Th\'eorique, CNRS, 
Luminy, Case 907\\ 
F-13288 Marseille Cedex 9 (France)}
\altaffiliation{
UMR 6207 du CNRS associ\'ee aux 
Universit\'es d'Aix-Marseille I et II et Universit\'e du Sud Toulon-Var; Laboratoire 
affili\'e \`a la FRUMAM-FR2291.}
\email{duval@cpt.univ-mrs.fr}

\author{Z.~Horv\'ath}

\affiliation{
Institute for Theoretical Physics, E\"otv\"os
University\\
P\'azm\'any P. s\'et\'any 1/A\\
H-1117 Budapest (Hungary)}

\email{
e-mail:zalanh@ludens.elte.hu}

\author{P.~A.~Horv\'athy}
\affiliation{
Laboratoire de Math\'ematiques et de Physique Th\'eorique, 
Universit\'e de Tours.
Parc de Grandmont. 
F-37200 Tours (France)}

\email{horvathy@lmpt.univ-tours.fr}

\date{\today}

\begin{abstract}
Mimicking the description of spinning particles in General Relativity, the Fermat Principle is extended to spinning 
photons. Linearization of the resulting Papapetrou-Souriau
type equations yields the  semiclassical model used recently
to derive the  ``Optical Hall Effect'' for polarized
light (alias the ``Optical Magnus  Effect'').
\end{abstract}


\maketitle

%FermatRapid.tex 

%%%%%%%%%%%%%%%%%%%%%%%%%%%%%%%%%%%%%%%%%%%%%%%%%%%%%%%%%%%%%%%%%%%%%
%%%%%%%%%%%%%%%%%%%%%%%%%%%%%%%%%%%%%%%%%%%%%%%%%%%%%%%%%%%%%%%%%%%%%
%\section{Introduction}
%%%%%%%%%%%%%%%%%%%%%%%%%%%%%%%%%%%%%%%%%%%%%%%%%%%%%%%%%%%%%%%%%%%%%
%%%%%%%%%%%%%%%%%%%%%%%%%%%%%%%%%%%%%%%%%%%%%%%%%%%%%%%%%%%%%%%%%%%%%

Light is an electromagnetic wave, whose propagation is described by 
Maxwell's theory. 
It can also be viewed, however, as a particle (a ``photon''). 
Here we adhere to the second approach: we describe
light by a {\it bona fide} mechanical model in that we use 
a Lagrangian.  

In traditional geometrical optics the spin degree of freedom is  
neglected, and  the light rays obey the Fermat Principle \cite{BW}.  
In the intermediate model advocated by Landau and Lifchitz \cite{LL8},
the photon is polarized, but the polarization is simply carried along
by the light rays, and has no influence on the trajectory of light.
Recent approaches \cite{OMN,BB} go one step further~: the
 feedback from the polarization deviates the trajectory
from that given by the Fermat Principle. A dramatic consequence is
that, for polarized light, the Snel(-Descartes) law  of refraction 
requires correction~: the plane 
 of the refracted (or reflected) ray is
 shifted perpendicularly to that of the incident ray \cite{OMN}. 
 This  ``Hall Effect for light''  is a manifestation of
the Magnus-type interaction 
between the refractive medium and the photon's polarization 
\cite{BB}.
It can be derived in a semiclassical framework, 
 which also includes a Berry-type term  \cite{Niu,AHE,SpinHall}.

In this Rapid Communication, we argue that the deviation of polarized light
from the trajectory predicted by ordinary geometrical optics is indeed analogous to the deviation 
of a spinning particle from geodesic motion in General Relativity.
The resulting equations 
  are  reminiscent of those of Papapetrou and Souriau
\cite{PapaSou}.

%%%%%%%%%%%%%%%%%%%%%%%%%%%%%%%%%%%%%%%%%%%%%%%%%%%%%%%%%%%%%%%%%%%%%%%%%%%
%\section{Fermat Principle for the spinning photon}\label{Fermatprinc}
%%%%%%%%%%%%%%%%%%%%%%%%%%%%%%%%%%%%%%%%%%%%%%%%%%%%%%%%%%%%%%%%%%%%%%%%%%%%%

In detail, the Fermat Principle of geometrical optics
says that light in an isotropic medium of refractive index 
$n=n(\br)$ propagates along curves 
that minimize the optical length. 
Light rays are hence geodesics of the ``optical'' metric $g_{ij}=n^2(\br)\delta_{ij}$ of 3-space.
 To extend this theory to spin we consider the 
bundle of positively oriented orthonormal frames over a 3-manifold endowed with a Riemannian metric $g_{ij}$. 
At each point,  such a ``Dreibein''
 is given by three  orthogonal vectors 
 $U^i,V^i,W^i$ of unit length that span unit volume.
We stress that the [$6$-dimensional] orthonormal frame bundle 
we are using here is a mere 
artifact that allows us to define a variational formalism.
Eliminating unphysical degrees of freedom will leave us with $4$ independent physical variables.

Introducing the covariant exterior derivative associated with the Levi-Civita connection,
$DU^k\!=\!dU^k\!+\!\Gamma^k_{ij}dx^iU^j$, we posit the 
reparametrization-invariant action  
\begin{equation} 
S=
S_\mathrm{Fermat}+S_{\mathrm{spin}}=\kappa\int U_i\frac{dx^i}{d\tau}
d\tau
-s\int V_i\frac{D W^i}{d\tau}d\tau,
\label{spinlag0}
\end{equation}
where $\tau$ is some parameter along the light ray. 
The parameters  $s$ and $\kappa>0$  are interpreted 
as the spin and the color, respectively.  Upon first quantization, 
$\kappa$ becomes indeed, for a monochromatic wave,
$2\pi\hbar/\lambda$, where~$\lambda$ is the wavelength \cite{SSD}.
For the photon $s=\pm\hbar$, but we  keep it arbitrary for future convenience. 
Equation~(\ref{spinlag0}) is supplemented with the constraints 
$
U_iU^i=V_iV^i=W_iW^i=1, 
$ and
$\
U_iV^i=U_iW^i=V_iW^i=0.
$ 

The first term in (\ref{spinlag0}) is [$\kappa$ times] the usual optical length; the second,
``Wess-Zumino-type" \cite{WZ} term, that arises naturally
in the geometric framework of \cite{SpinOptics},
corresponds to the Berry connection, and is indeed analogous to the
torsion term considered by Polyakov \cite{tors}. 

The Euler-Lagrange equations  are obtained as follows.
%Let $x^i(\tau)$ be a curve in space, and let $\delta x^i(\tau)$ be its variation. 
Variation of the first term in (\ref{spinlag0}) yields
\begin{equation}
\delta S_\mathrm{Fermat}=\kappa\int\left[-\delta x_k\frac{DU^k}{d\tau}+\frac{dx^k}{d\tau}
\delta_\Gamma U_k\right]d\tau,
\end{equation}
where $\delta_\Gamma  U^k=\delta U^k+\Gamma^k_{lm}\delta x^l U^m$
is the covariant variation of the vector field $U^i$.
For the spin term, straightforward calculation yields
\begin{equation}
\delta S_{\mathrm{spin}}=\int\left[
\delta_{\Gamma}U_j\frac{DU_k}{d\tau}
+
\frac{1}{2}\delta x^i\frac{dx^\ell}{d\tau}R_{jki\ell}
\right]S^{jk} d\tau,
\end{equation}
where $R_{jki\ell}=g_{im}(\p_j\Gamma^m_{k\ell}-\p_k\Gamma^m_{j\ell}+\cdots)$
 is the Riemann tensor and  $S^{ij}=-s(V^iW^j-W^iV^j)$ is the spin tensor. 
 Then the variational principle $\delta S=0$  allows us to infer the pair of equations
\begin{eqnarray}
&&\kappa g_{ik}\displaystyle\frac{dx^k}{d\tau}+S_{ik}
\displaystyle\frac{DU^k}{d\tau}=\mu U_i,
\label{xeq}
\\[8pt]
&&\kappa g_{ik}\displaystyle\frac{DU^k}{d\tau}-
\displaystyle\frac{1}{2}R_{jki\ell}S^{jk}\displaystyle\frac{dx^\ell}{d\tau}=0,
\label{ueq}
\end{eqnarray}
where  $\mu$ is a Lagrange multiplier which enforces
the orthogonality of $U^i$ and $\delta_{\Gamma}U^i$.
Inserting ${DU^k}/{d\tau}$  into Eqn~(\ref{xeq})
and redefining the parameter along the ray by $dt= (\mu/\kappa)d\tau$
 yields the Papapetrou-Souriau type \cite{PapaSou} equations, 
\begin{eqnarray}
\frac{dx^i}{dt} 
&=&U^i+\frac{1}{2}\frac{S^i_{\ j}R(S)^j_{\ k}U^k}
{(\kappa^2+s^2E_{\ell m}U^\ell U^m)},
\label{genvelrel}
\\[6pt]
s\frac{DU^i}{dt}&=&-\frac{1}{2\kappa }
R(S)^i_{\ j}\frac{dx^j}{dt},
\label{genLorentz}
\end{eqnarray}
where  
$E_{ij}=R_{ij}-\frac{1}{2}Rg_{ij}$ is the Einstein tensor
 and the matrix
$R(S)^{\ell}_{\ k}= R^{\ell}_{ijk}S^{ij}$ represents the interaction 
of spin with the curvature, responsible for tidal forces.
 Due to the spin-curvature coupling,
 the direction of the velocity 
differs, in general, from that of the spin vector  $sU^i=-\half\sqrt{g}\epsilon^{ijk}S_{jk}$.
 
For the optical metric, the Christoffel symbols are
$\Gamma^k_{ij}=\frac{1}{n}\big(\p_in\,\delta^k_j+\p_jn\,\delta^k_i-\p^kn\,\delta_{ij}\big)$. (The optical metric
 is hence not flat unless the refraction index is constant.)
Putting $\br=(x^i)$, $u^i=nU^i, v^i=nV^i, w^i=nW^i$, 
introducing the momentum,
\begin{equation}
\bp=n\Big[\kappa\bu+s\, \grad\big(\frac{1}{n}\big)\times\bu\Big],
\label{momentum}
\end{equation}
and denoting the derivative w.r.t. 
$t$ by a ``dot'', 
our  Lagrangian in (\ref{spinlag0}) can also be presented  as 
$L=
\bp\cdot\dot{\br}-s\bv\cdot\dot{\bw}$.
The equations of motion for $\br$ and $\bp$ read, in this case,
\begin{eqnarray}
\dot{\br}&=&aA\bp+
\frac{s^2}{n\kappa^2}\,\grad\big(\grad(\frac{1}{n})\big){A\bp},
\label{horror1}
\\[8pt]
\dot{\bp}&=&
-n(\bp\cdot\dot{\br})\,\grad(\frac{1}{n})\nonumber
\\[3pt]
&&+\frac{s}{\kappa}\,\grad\big(\grad(\frac{1}{n})\big){A\bp}\times\dot{\br},
\label{horror2}
\end{eqnarray}
where
$a=1+(s^2/\kappa^2)\,\big((\grad(1/n))^2-(1/n)\Delta(1/n)\big)$
 and
\begin{eqnarray}
&A\bp=\displaystyle\frac{1}{1+(s/\kappa)^2\,\big(\grad(\frac{1}{n})\big)^2}
\times\nonumber
\\[4pt]
&\left[
\bp-\displaystyle\frac{s}{\kappa}\,\grad(\frac{1}{n})\times\bp
+\displaystyle\frac{s^2}{\kappa^2}\,\big(\grad(\frac{1}{n})\cdot\bp)\,
\grad(\frac{1}{n})
\right].
\end{eqnarray} 
These equations describe spinning light in an inhomogeneous medium. Let us mention, for completeness, that the evolution of
the spin vector, which follows from  (\ref{genLorentz}),
is given by
$ 
s\dot{\bu}=-n\kappa\big[\dot{\br}-\frac{s}{\kappa}\,\grad(\frac{1}{n})\times\dot{\br}\big] \times\bu.
$ 

If the medium is spherically symmetric, $n=n(r)$, conserved angular momentum is readily derived using Noether's theorem.
It reads
\begin{equation}
\bJ=\br\times\bp+s\bu.
\label{genangmom}
\end{equation}

Let us now discuss some particular cases of our general theory.

(i) For $s=0$ we have $\bp^2=n^2\kappa^2$,
$a=1$ and ${A\bp}=\bp$. Introducing the elementary arc length
$d\sigma=n\kappa dt$, we recognize the usual Fermat equations, 
$ n{d\br}/{d\sigma}=\bp/\kappa,$ 
${d(\bp/\kappa)}/{d\sigma}=\grad n$ \cite{BW}.  

 (ii) In a homogeneous medium, $n=\const.$ we get, for any
color, $\kappa$, and spin, $s$, 
the same equations: light propagates along straight
lines parallel to $\bp=n\kappa\bu$. 
The  model is invariant w.r.t.
the Euclidean group $\mathrm{SE}(3)$ consisting of space translations and rotations.
The associated conserved quantities are the linear momentum,
$\bp=n\kappa\bu$, and the angular momentum,
(\ref{genangmom}), which is now $\bJ=\br\times\bp+s\bp/p$.

(iii) In a medium with slowly varying refractive index, terms involving second-order derivatives and quadratic expressions in $\grad(1/n)$ can be neglected, e.g., $R_{ij}=\frac{2\p_in\p_jn}{n^2}-\frac{\p_i\p_jn}{n}
-\frac{\Delta n}{n}\delta_{ij}\approx0$. Hence the trajectory of light is approximately tangent to the spin and the latter is
approximately parallel transported,
\begin{equation}
\frac{dx^i}{d\tau} 
\approx U^i,
\quad
\frac{DU^i}{d\tau}\approx0.
\label{lineq}
\end{equation}
In $\bp$-terms, 
$\bp^2\approx n^2\kappa^2$,
 and the general equations (\ref{horror1})-(\ref{horror2}) are approximated by
\begin{eqnarray}
\dot{\br}\approx\bp-\frac{s}{\kappa}\,\grad(\frac{1}{n})\times\bp,
\quad
\dot{\bp}\approx 
-n^3\kappa^2\grad(\frac{1}{n}).
\label{LinDuv}
\end{eqnarray}
In the case of spherical symmetry, the general angular momentum (\ref{genangmom})
reduces, up to the approximately conserved extra term 
$(s^2/\kappa)\grad(1/n)\times\bu$,
 to the expression used by Onoda et al. in \cite{OMN}, namely to
\begin{equation}
\bJ^{\mathrm{OMN}}=\br\times\bp+s\frac{\bp}{p}.
\label{OMNangmom}
\end{equation}

 Let first consider the free case, $n=1$.
The variable~$\br$ used so far has been an arbitrary point of
the light ray. Now, the ray itself can be labeled by its direction 
$\bu$ and
$ 
\bq=\br-(\bu\cdot\br)\,\bu,
$ 
which is in fact the shortest
vector drawn from the origin to the ray (orthogonal to 
the unit vector~$\bu$), 
and can be thought of as the ``position'' of the ray.
The $4$-dimensional manifold, ${\cal M}$, of light rays described by 
$\bu$ and
$\bq$, 
has the topology of the tangent bundle of the two-sphere and can be 
identified with a coadjoint orbit of $\mathrm{SE}(3)$. 
The Casimir invariants of the orbit
(which determine unitary irreducible representations) are 
$\kappa=p\ (=\sqrt{\bp^2})$
and
$s=\bJ\cdot\bp/p$. 
The corresponding orbit, ${\cal M}$, is endowed with the canonical symplectic structure
\begin{equation} 
\omega_0=\kappa du_i\wedge dq^i-\frac{s}{2}\epsilon_{ijk}u^idu^j\wedge du^k,
\label{freeO}
\end{equation} 
see  \cite{SSD}.
 The  monopole-like term in (\ref{freeO}) is the Berry curvature, 
 $\frac{1}{2}s\epsilon_{ijk}p^idp^j\wedge dp^k/p^3$. It
makes the components of $\bq$ non-commuting \cite{SpinOptics}, 
i.e., Cartesian coordinates $q_1$ and $q_2$
have non-vanishing Poisson bracket,
$
\big\{q_1,q_2\big\}={s}/{\kappa^2}.
$
Upon (first) quantization, in the case $s=\hbar$, the quantum ``position'' operators $\hat{q}_1$ and $\hat{q}_2$ satisfy
\begin{equation} 
\big[\hat{q}_1,\hat{q}_2]=i({\lambda}/{2\pi})^2,
\label{qHeisenberg}
\end{equation}
where $\lambda=2\pi\hbar/\kappa$. The Heisenberg uncertainty relation read, therefore,
$ 
\Delta\hat{q}_1\cdot\Delta\hat{q}_2\geq\frac{1}{2}({\lambda}/{2\pi})^2,
$ 
which provide a new interpretation of the localization defect of spinning light rays, limiting the resolving power of optical instruments to the order of the wavelength,
$\Delta\hat{q}\approx{}O(\lambda)$.

In a non-trivial refractive medium the (exact) two-form (\ref{freeO}) is replaced by
\begin{eqnarray}
\omega&=&\kappa DU_i\wedge dx^i
-\frac{1}{4}R(S)_{ij}dx^i\wedge dx^j\nonumber
\\
&&-\frac{s}{2}\sqrt{g}\,\epsilon_{ijk}U^iDU^j\wedge DU^k
\label{symplectic}
\end{eqnarray}
on the orthonormal frame bundle. 
The Euler-Lagrange equations (\ref{genvelrel})-(\ref{genLorentz}) 
correspond in fact to the kernel of the two-form (\ref{symplectic}),
 see \cite{SpinOptics}.
Conversely, the spin term in our Lagrangian  
comes  from a potential for the spin terms in the two-form
(\ref{symplectic}).

%%%%%%%%%%%%%%%%%%%%%%%%%%%%%%%%%%%%%%%%%%%%%%%%%%%%%%%%%%%%%%%%%%%%%%%%%%%%
%\section{Relation to previous work}
%%%%%%%%%%%%%%%%%%%%%%%%%%%%%%%%%%%%%%%%%%%%%%%%%%%%%%%%%%%%%%%%%%%%%%%%%%%%

Now  putting $\bp\to\bp/\kappa$ and  $\tau\to\kappa\tau$,
our linearized equations (\ref{LinDuv}) become those
proposed in \cite{BB}. 

 The relation to the model of Onoda et al.  \cite{OMN} is more subtle.
In their approach, polarization is
an additional variable,  represented by a two-component 
complex vector, $z=(z_a)$ with $a=\pm$, such that $|z_+|^2+|z_-|^2=1$, 
acted upon by $\su(2)$. 
Their semi\-classical equations of motion can be written as
\begin{equation}
{\br}'=\frac{1}{nk}\bk+{\bk}'\times{\bf\Omega}^{ab}\bar{z}_a z_b,
\quad
{\bk}'=-\grad(\frac{1}{n})k,
\label{OMN1-2}
\end{equation} 
supplemented with
\begin{equation}
z'_a=k\,\grad(\frac{1}{n})\cdot{\bf\Lambda}^{ab}\,z_b,
\label{OMN3}
\end{equation}   
where  $\bk$ is the wave vector and ${\bf\Lambda}^{ab}(\bk)$
is an $\su(2)$-valued non-abelian ``Berry'' vector potential. The Berry curvature can be represented by an $\su(2)$-valued vector
${\bf\Omega}=\sigma_3\bk/k^3$, %which is in fact 
a Dirac monopole in ${\bf k}$-space, diagonally embedded
into $\su(2)$.
The vector potential  for ${\bf\Omega}$ can, 
therefore, be chosen as
${\bf\Lambda}^{ab}=i{\bf\Lambda}(\sigma_3)^{ab}$
where ${\bf\Lambda}(\bk)$ is a  monopole potential,
 $\rot\,{\bf\Lambda}=\bk/k^3$. 
The number of equations in (\ref{OMN1-2})-(\ref{OMN3}) can be
reduced to two. The
polarization equation (\ref{OMN3}) can in fact be 
solved  formally by parallel transport,
$z_a=e^{ia\theta}z_a^0$,  where the phase is given by the non-integrable phase factor
$\theta=\int\!k\,\big(\grad(\frac{1}{n})\cdot{\bf\Lambda}\big)d\sigma$. 
Then the Berry term
becomes simply  ${\bf\Omega}^{ab}\bar{z}_az_b=s\bk/k^3$ where 
 $s=|z_+|^2-|z_-|^2=|z_+^0|^2-|z_-^0|^2$, since the $|z_a|^2$ 
are separately conserved.
Notice that this $s$ is a constant of the motion,
which can take any value between $-1$ and $+1$.
Identifying the wave vector, ${\bf k}$, with our
momentum, $\bp$, and putting
$(\,\cdot\,)'=(n^2\kappa)d/dt$ 
transforms finally (\ref{OMN1-2})
into our equations (\ref{LinDuv}).

%%%%%%%%%%%%%%%%%%%%%%%%%%%%%%%%%%%%%%%%%%%%%%%%%%%%%%%%%%%%%%%%%%%%%%%%%%%%
%\subsection{Polarization} 
%%%%%%%%%%%%%%%%%%%%%%%%%%%%%%%%%%%%%%%%%%%%%%%%%%%%%%%%%%%%%%%%%%%%%%%%%%%%

We did not consider the polarization in our framework. 
As long as we are only interested in
describing light rays, polarization is a secondary quantity, 
whose only
role is to generate spin, which in turn deviates the trajectory from that of conventional geometrical optics. 
It is hence more appropriate to speak of
spinning light than of polarized light.
 Let us nevertheless mention that first quantization 
 along the lines of \cite{SSD}
 of the classical model (\ref{freeO}), with Casimirs $\kappa$ and $s=\pm\hbar$, yields, in the gauge $\mathrm{div}\bA=0$,
the vectorial Helmholtz equation 
\begin{equation}
(\Delta+k^2)\bA=0,
\label{Helm}
\end{equation} 
where $k=\kappa/\hbar$. It follows that
$\bE=ik\bA$ and $\bB=\rot\,\bA$
satisfy the field equations
\begin{eqnarray}
\label{rotE}
\rot\,\bE-ik\bB&=&0,
\\
\label{rotB}
\rot\,\bB+ik\bE&=&0,
\end{eqnarray}
associated with our Euclidean model. 
Promoting the parameter $t$  as ``time'', these 
are indeed the vacuum Maxwell equations for
the stationary fields $\bE e^{-ik t}$ and $\bB e^{-ik t}$,
respectively ($c=1$).
In an isotropic medium, 
Eqn (\ref{rotB}) is generalized \cite{BW} to
\begin{eqnarray}
\label{nrotB}
\rot\,\bH+ik\bD&=&0,
\end{eqnarray}
where $\bD=\epsilon\bE$ and $\bB=\mu\bH$, with
$\epsilon$ and $\mu$ the permittivity and permeability,
respectively.
The refractive index is $n=\sqrt{\epsilon\mu}$. Remarkably, the field equations can, again,
be rewritten in terms of ``optical'' metric, namely in the form
(\ref{rotE}) and (\ref{nrotB})
above, replacing the operator $\rot$ by its curved-space
form, $\rot\,{\bf X}\to n^{-3}\rot(n^2{\bf X})$, and
rescaling the fields,
$\bE\to n^{-2}\bE$, $\bH\to n^{-2}\bH$,
$\bB\to n^{-3}\bB$, $\bD\to n^{-3}\bD$
and 
$\epsilon\to n^{-1}\epsilon,\, \mu\to n^{-1}\mu$.

Conventional geometric optics can be derived from the
 eikonal approximation of Maxwell's electrodynamics \cite{BW}. 
 Here we followed the
opposite way~: we started with a classical model and
derived the stationary Maxwell equations (\ref{rotE}-\ref{nrotB})
by (first) quantization. Although we have not yet been able to
deduce our action (\ref{spinlag0})  from taking a suitable semiclassical
 limit of (\ref{rotE}-\ref{nrotB}),
we emphasise that our model
actually comes from \textit{first principles} -- but of those of Mechanics. \cite{SpinOptics}.
Firstly, the  free model is constructed along the lines
suggested by Souriau's  \cite{SSD}, applied to the Euclidean group.  The second step is
\textit{minimal gravitational coupling}, 
which amounts  to replacing the ordinary scalar product by 
the one associated with the ``optical'' metric 
and the ordinary derivatives by  covariant 
derivatives; this yields the two-form (\ref{symplectic}).
The latter is in turn associated with
 first-order variational calculus on
``phase space'', whose Lagrangian is precisely our
 (\ref{spinlag0}).
 
Our model reproduces, at first order, the phenomenological
descriptions proposed in \cite{OMN, BB} which can, in turn,
be derived by taking an improved semiclassical limit
of the Maxwell equations \cite{BB}. Does a similar
procedure work for our model~? The question is open.  
 
Our theory is neither relativistic nor non-relativistic,
since it does not involve time at all; it is  based on the
Euclidean group -- which is indeed a subgroup
of both the Galilei and of the Poincar\'e groups. 
Our Euclidean model arises in fact
as a \textit{reduction} by time translations
of  the zero-mass spinning orbits of \textit{both
 the Galilei and the Poincar\'e groups} \cite{SpinOptics} 
 -- which constitute the conventional descriptions 
 of ``classical light'' \cite{SSD,WZ}).

An application of our semiclassical model is the derivation of
the modified laws of refraction and reflection at the interface of
two homogeneous media with different refractive indices.
As  found by Onoda et al. \cite{OMN}, polarized light
suffers, in fact, a transverse shift.
This  ``Optical Hall effect'' \cite{OMN}
is indeed an optical version of the
spin-Hall effect \cite{SpinHall}.
Their shift formula  can be rederived \cite{SpinOptics},
following Souriau \cite{SSD}, who argues that
the two ``mechanical'' states  on both sides of the interface
are related by a symplectic transformation,  $S$, which is
indeed the classical counterpart of the quantum
scattering matrix. This transformation commutes 
with the symmetries of the optical device; in our case,
this is plainly the Euclidean group generated
by translations of the separating plane and  
rotations around its normal direction,~$\bN$.  
Tedious  calculation provides us with
 the ``classical scattering matrix'', $S$ \cite{SpinOptics}. 

Firstly, the conservation of planar linear
momentum extends Snel's laws  to spinning light, namely
\begin{equation}
n_\mathrm{in}\sin\theta_\mathrm{in}=n_\mathrm{out}\sin\theta_\mathrm{out},
\label{Snel1}
\end{equation}
for refraction, and 
$
\theta_\mathrm{in}=\pi-\theta_\mathrm{out},
$ for reflection, respectively,
where $n_\mathrm{in}$ resp.d $n_\mathrm{out}$
denote the refractive indices on both sides of the interface.

\goodbreak

Next, the conservation of the (planar) angular momentum implies
that light is shifted transversally \cite{OMN, SpinOptics}, viz. 
\begin{equation}
\bq_\mathrm{out}-\bq_\mathrm{in}
=
\frac{\left[
s_\mathrm{out}\cos\theta_\mathrm{out}-s_\mathrm{in}\cos\theta_\mathrm{in}
\right]}
{\kappa\,n_\mathrm{in}|\sin\theta_\mathrm{in}|}\,\frac{\bN\times\bu_\mathrm{in}}
{|\bN\times\bu_\mathrm{in}|},
\label{shift}
\end{equation} 
where $s_\mathrm{in}=s_\mathrm{out}$ for refraction. 

Notice that the shift depends in general on the wavelength,  
$\lambda=2\pi\hbar/\kappa$, white light is split, 
in general, into colors, shifted by different amounts.
For $n_\mathrm{out}=-n_\mathrm{in}$, however, Snel's laws entail that the shift  (\ref{shift}) \textit{vanishes}: 
white light is \textit{not} decomposed
and is indeed refracted
following the classic Snel law, as if it had no spin~! 

This case is \textit{not} of pure academic interest, 
owing to the existence of left-handed media (with a negative refractive 
index) \cite{negind}. 
In the ideal case, one can have $n=-1$, and a simple slab
with parallel sides  \cite{Pendry} provides us with a ``perfect
lens'' with no chromatic aberration.

We note that the shift (\ref{shift})  vanishes also for a reflection,  since then
$s_\mathrm{in}=-s_\mathrm{out}$. This does not contradict the
results in Imbert \cite{Imbert}, which are indeed of higher-order.

\goodbreak

\begin{acknowledgments}
We are indebted to Prof. M. Berry and to Dr. K. Bliokh for correspondence.
\end{acknowledgments}

%%%%%%%%%%%%%%%%%%%%%%%%%%%%%%%%%%%%%%%%%%%%%%%%%%%%%%%%%%%%%%%%%%%%
%%%%%%%%%%%%%%%%%%%%%%%%%%%%%%%%%%%%%%%%%%%%%%%%%%%%%%%%%%%%%%%%%%%%


\begin{thebibliography}{99}
%%%%%%%%%%%%%%%%%%%%%%%%%%%%%%%%%%%%%%%%%%%%%%%%%%%%%%%%%%%%%%%%%%%%
%%%%%%%%%%%%%%%%%%%%%%%%%%%%%%%%%%%%%%%%%%%%%%%%%%%%%%%%%%%%%%%%%%%%

\bibitem{BW}
 M. Born and E. Wolf,
 {\it Principles of optics}. 7th ed.
  Cambridge UP (1999). 

\bibitem{LL8}
L. Landau, E. Lifchitz,
{\it \'Electrodynamique des milieux continus}.
Cours de Physique Th\'eorique vol. 8. Moscou: MIR (1969).
 
\bibitem{OMN}
M. Onoda, S. Murakami, and N. Nagaosa,
%{\it Hall effect for light}.
 {\sl Phys. Rev. Lett}. {\bf 93}, 083901 (2004)
 [\texttt{cond-mat/0405129}];
 M. Onoda,
% {\it Spinning waves and the optical Hall Effect}.
IEEE LEOS NEWSLETTER, {\bf 18}, 5, dec. 2004.

\bibitem{BB}
 V. S. Liberman and B. Ya. Zeldovich,
% {\it Spin-orbit interaction of a photon \dots}
 {\sl Phys. Rev.} {\bf A 46}, 5199 (1992);
 K. Yu. Bliokh, Yu. P. Bliokh,
%{\it Topological spin transport of photons: the optical Magnus effect and Berry Phase \dots}.
 {\sl Phys. Lett.} {\bf A333}, 181 (2004)
[\texttt{physics/0402110}],
K. Y. Bliokh and Y. P. Bliokh, 
%{\it Modified geometrical optics of a smoothly inhomogeneous isotropic medium: the anisotropy, Berry phase, and the optical Magnus effect.} 
{\sl Phys. Rev.} {\bf E 70}, 026605, (2004),
[\texttt{physics/0402014}];
K. Y. Bliokh and Y. P. Bliokh, 
%{\it Conservation of Angular Momentum, Transverse Shift, and Spin Hall Effect in %Reflection and Refraction of Electromagnetic Wave Packet}, 
[\texttt{physics/0508093}].
 
\bibitem{Niu}
A. Bohm, A. Mostafazadeh, H. Koizumi,
Q. Niu and J. Zwanziger,
{\it The Geometric Phase in Quantum Systems}.
Chapter 12. Springer Verlag (2003).

\bibitem{AHE}
T. Jungwirth, Q. Niu, and A. H. MacDonald,
{\sl Phys. Rev. Lett.} {\bf 88}, 207208 (2002);
Fang et al. {\sl Science} {\bf 302}, 92 (2003).

\bibitem{SpinHall}
S. Murakami, N. Nagaosa, and S.-C. Zhang,
{\sl Science}  {\bf 301}, 1348 (2003)
[\texttt{cond-mat/0308167}].

\bibitem{PapaSou}
A. Papapetrou, {\sl Proc. Roy. Soc.} {\bf A 209}, 248 (1951);
%{\it Spinning test particles n General Relativity.1.}.
 J.-M.~Souriau,
%{\it Mod\`ele de particule \`a spin \dots}
% dans le champ % %\'electromagn\'etique  et gravitationnel},
 {\sl Ann. Inst. Henri Poincar\'e}, {\bf 20 A}, 315 (1974).

\bibitem{SSD}
J.-M.~Souriau,
{\it Structure des syst\`emes dynamiques},
Dunod: Paris (1970). [English edition:
{\it Structure of Dynamical Systems: a Symplectic View of Physics}.
Birkh\"auser~(1997).] 
 
 
\bibitem{WZ}
B.-S. Skagerstam,
%{\it Localization of massless spinning particles and the Berry phase}.
 [\texttt{hep-th/9210054}]; 
 A. P. Balachandran, G. Marmo, A. Simoni, A. Stern, F. Zaccaria:
%{\it On a classical description of massless particles}.
Proc. ISATQP-Shanxi Conf. (1992)
pp. 396-402. Ed. J-Q. Liang, M.L. Wang, S.N.Qiao, D.C.Su
Science Press Beijing (1993); 
A. B\'erard and H. Mohrbach,
 {\sl Phys. Lett.} {\bf A352}, 190 (2006), 
 [\texttt{hep-th/0404165}];
P. Gosselin, A. B\'erard and H. Mohrbach, [\texttt{hep-th/0603227}].

 
\bibitem{SpinOptics}
 C. Duval, Z. Horv\'ath, P. A. Horv\'athy,
% {\it Geometrical SpinOptics and the Optical Hall Effect}.
 [\texttt{math-ph/0509031}].
 

\bibitem{tors}
 A. M. Polyakov,  {\sl Mod. Phys. Lett.} {\bf A 3} 325 (1988);
M. Plyushchay, {\sl Phys. Lett.} {\bf B 262}, 71 (1991);
 {\sl Nucl. Phys.} {\bf B362}, 54 (1991), etc.
 
\bibitem{negind}
V. G. Veselago, {\sl Sov. Phys. Solid State} {\bf 8}, 3571 (1966); 
{\sl Sov. Phys. Usp}. {\bf 10}, 509 (1968).
See K. Yu. Bliokh and Yu. P. Bliokh,
{\sl Uspekhi} {\bf 47}, 393 (2004)
for a comprehensive review.

\bibitem{Pendry}
J. B. Pendry,
{\sl Phys. Rev. Lett.}
{\bf 85}, 3966 (2000).

\bibitem{Imbert}
C. Imbert, 
%{\it Calculation an Experimental Proof of the Transverse Shift Induced by Total 
%Internal Reflection of Circularly Polarized Light Beam.},
 {\sl Phys. Rev.} {\bf D5}, 787 (1972).
  
\end{thebibliography}
\end{document}